# Planetary boundary layer depth in Global climate models induced biases in surface climatology

Richard Davy[1] & Igor Esau[1]

**The Earth has warmed in the last century with the most rapid warming occurring near the surface in the arctic[1,2,3]. This enhanced surface warming in the Arctic is partly because the extra heat is trapped in a thin layer of air near the surface due to the persistent stable-stratification found in this region[4]. The warming of the surface air due to the extra heat depends upon the amount of turbulent mixing in the atmosphere[5], which is described by the depth of the atmospheric boundary layer (ABL). In this way the depth of the ABL determines the effective response of the surface air temperature to perturbations in the climate forcing[6]. The ABL depth can vary from tens of meters to a few kilometers which presents a challenge for global climate models which cannot resolve the shallower layers[7]. Here we show that the uncertainties in the depth of the ABL can explain up to 60% of the difference between the simulated and observed surface air temperature trends and 50% of the difference in temperature variability for the Climate Model Intercomparison Project Phase 5 (CMIP5) ensemble mean. Previously the difference between observed and modeled temperature was thought to be largely due to differences in individual model's treatment of large-scale circulation and other factors related to the forcing, such as sea-ice extent[8,9,10]. While this can be an important source of uncertainty in climate projections, our results show that it is the representation of the ABL in these models which is the main reason global**

[1]Nansen Environmental and Remote Sensing Center, Thormøhlensgt. 47, 5006 Bergen, Norway.

**climate models cannot reproduce the observed spatial and temporal pattern of climate change. This highlights the need for a better description of the stably-stratified ABL in global climate models in order to constrain the current uncertainty in climate variability and projections of climate change in the surface layer.**

The surface air temperature has become one of the most commonly-used metrics to assess our climate and climate change. This is partly because it is such a readily and widely observed measure of the climate system (and so makes for a robust metric for climate models) but also because it is a very important parameter in the anthroposphere. A proper understanding of how the surface air temperature varies is essential to our understanding of Earth's climate and how it responds to forcing. The magnitude of the surface air temperature response to forcing is determined by three components: the magnitude of the forcing, any feedback processes involved, and the effective heat capacity of the system[11]. The effective heat capacity of the atmosphere is defined by the region of turbulent mixing through which the heat is mixed i.e. it is defined by the depth of the ABL[5,6]. So we can define the near-surface temperature response to forcing through an energy budget model of the form:

$$\frac{d\theta}{dt} = \frac{Q}{\rho c_p h} \qquad (1)$$

where $Q$ [W m$^{-2}$] is the heat flux divergence within the ABL, $h$ [m] is the depth of the ABL, $\rho$ [kg m$^{-3}$] is the air density, $c_p$ [J kg$^{-1}$ K$^{-1}$] the heat capacity at constant pressure, and $\theta$ [K] is the potential temperature. Note that this is a reasonable approximation for well-mixed layers, where the potential temperature is constant with height, but may be more complicated in stable boundary layers where the potential temperature increases with height. An assessment of the response of stable boundary layers to increased radiative forcing using a single column model with a well-developed radiation scheme determined that one may expect an enhanced warming near the surface and a cooling aloft[12]. This has been ascribed to the increased turbulence within the ABL leading to better mixing, and so a weaker gradient in potential temperature with height. So a part of the signal of enhanced warming in stable boundary-layers may be due to the redistribution, rather than the accumulation, of heat within the ABL. While this highlights the complex relationship between radiative forcing and temperature response in stable boundary layers, we can still expect that the magnitude of the SAT response to forcing will be modulated by the depth of the boundary-layer.

One important factor in the choice of a metric by which to assess the performance of a climate model is the availability and reliability of an equivalent observational dataset. This is a particular challenge when we wish to assess the performance of climate models with respect to the conditions of the ABL since establishing the climatology of the ABL from observations has proved difficult. A comparison of the climatology of the ABL depth (as defined from the bulk Richardson number) between Radiosonde observations, reanalysis and global climate models over Europe and the continental US found that there were large uncertainties in the depth of the ABL: up to 50% for shallow, stable boundary layers, and around 20% for the deeper, convective boundary-layer[13,14]. There is also a general bias in the models towards deeper boundary-layers, as the models have difficulties representing stably-stratified conditions. The ABL depth can even be lower than the first atmospheric model level of a global climate model. With this bias towards deeper layers we would expect, from equation 1, that the temperature response to forcing is under-estimated in models under conditions of stable-stratification, resulting in an under-estimation of the temperature variability. Given the very different description of the ABL depth between GCMs in stably-stratified conditions, we may also expect the models to have the greatest uncertainty as to the temperature trends and variability under such conditions. We have tested these hypotheses by assessing the performance of global climate models for metrics that are directly affected by the ABL-response mechanism (equation 1): the SAT mean, trends and variability, under different conditions of the boundary layer. These were used to assess model skill with respect to a reanalysis product, ERA-Interim[15].

We have used a statistical measure of model fidelity[16] to determine model departure from reanalysis for the historical simulations of the CMIP5 program, over the period 1979 - 2005. This is the full period of overlap between the CMIP5 historical runs and the ERA-Interim reanalysis dataset, which we use as our reference set. We also assess the performance of these models as an ensemble by determining the inter-model mean and spread of these metrics. The greatest model departure from observations is seen for shallow, stably-stratified boundary layers. This is also where we see the greatest inter-model spread in our metrics. In this work we have focused on the northern hemisphere because we use an approach which considers model error over a seasonal cycle and that (given the importance of cold, stably-stratified conditions that this work highlights) these results are of principal interest for high-latitude, continental-interior regions. Nevertheless we obtain very similar results from a global analysis due to our approach which allows us to look at model performance with regard to the physics independently of the frequency of occurrence of a given state. So while the inclusion of the tropics in our analysis will add more cases of convective conditions, it does not alter our results for model error under stable-stratification.

**Model biases and error as a function of ABL conditions**

The biases in the simulated SAT mean, trends and variability over land, with respect to ERA-Interim, are given in Figure 1. They are plotted as a function of surface sensible heat flux. All model results approach a minimum bias in the mean SAT for weakly convective conditions which generally increases as we move towards more strongly convective conditions. But the largest biases are seen in the shallow, stably-stratified ABLs. There is a general warm bias in shallow

ABLs with mean-model biases in excess of 10 K in very shallow layers. The ACCESS 1.3 model is the only model which shows a strong bias in the mean SAT for weakly convective conditions: surface sensible heat flux in the range 20-70 $Wm^{-2}$. It is worth noting that this model has included many new developments to the core ACCESS model which have not yet been fully tested for long-period climate runs[32].

All models show a consistent bias in the trend and variability in SAT over all convective conditions. If a model over-estimates the trend in weak convective conditions, it tends to over-estimates the trend in strong convective conditions, and vice versa. However, in stable conditions we see larger biases in the trends and variability, with a general negative bias in the models. This pattern is most apparent in the model biases in SAT variability: there are small, consistent biases in convective conditions which rapidly become increasingly negative as we move towards increasingly stable stratification. This is not surprising since a linear temperature trend picks out a single mode of variability, the nature of which can be very sensitive to the period of investigation, whereas temperature variance is an integrated measure of all modes of variability in the period under analysis, and so will give a clearer measure of any systematic biases that affect the temperature response.

This under-estimation of the magnitude of SAT trends and variability in stable conditions is consistent with our expectation from equation 1. Given that models are biased towards deeper ABLs under stable stratification, and that these shallow layers can strongly affect the magnitude of the SAT response to forcing, we expect a positive bias in ABL depth to result in a negative bias in SAT trends and variability.

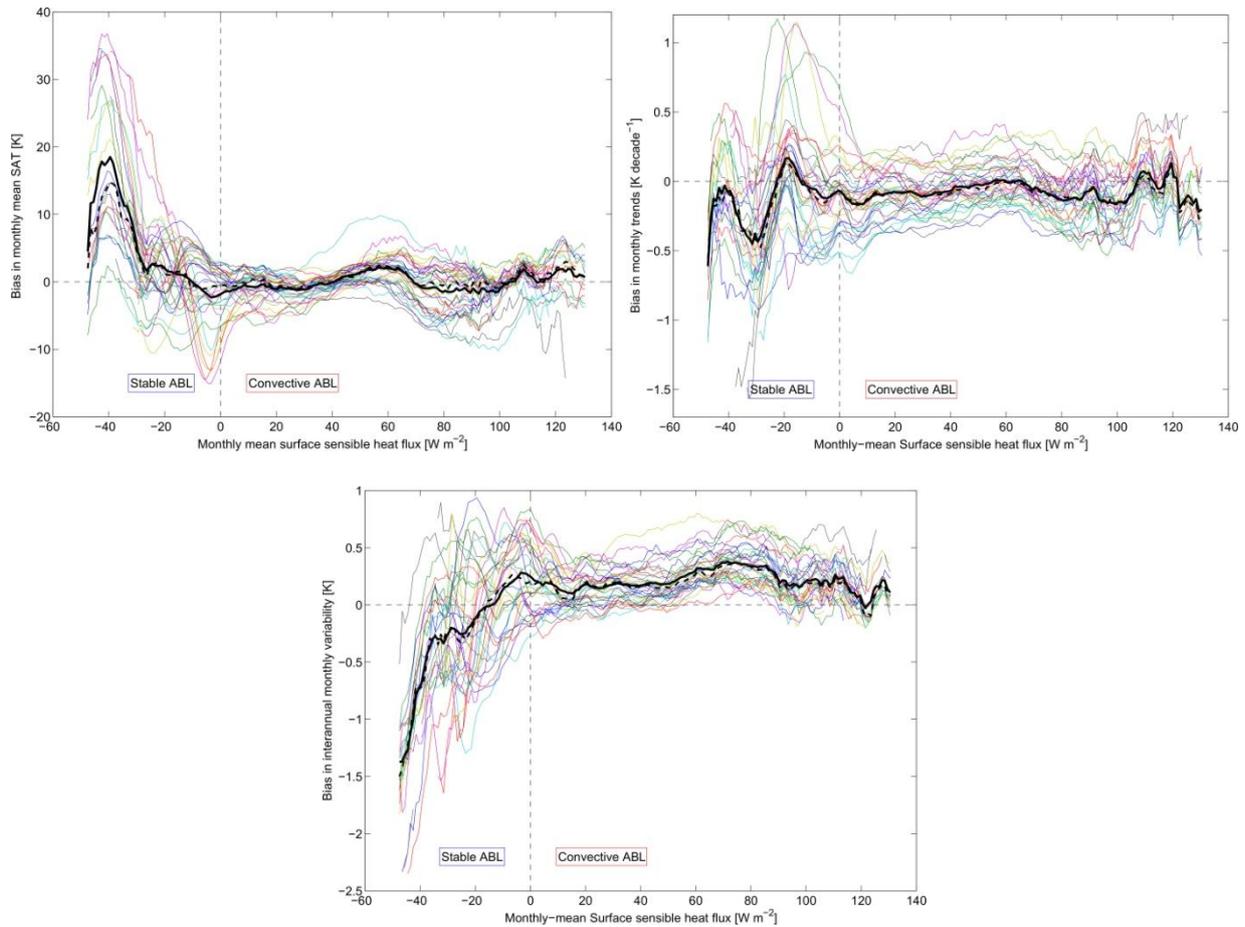

**Figure 1.** The bias (CMIP5–ERA Interim) in the mean, trend and standard deviation in surface air temperature, as a function of the mean surface sensible heat flux for locations over land. The colored lines represent 36 of the model results from the CMIP5 simulations, and the multi-model mean and median are highlighted in solid and dashed black lines respectively.

The RMS error of the SAT mean, trend and inter-annual variability reflects the pattern of model biases – large biases lead to large RMS errors. These are given as a function of surface sensible heat flux for locations over land, and as a function of lower tropospheric temperature stability over ocean (Figure 2). The models show the largest departure from the observed temperature trends and variability in shallow ABLs whether over land or ocean. The model error has a similar dependency on ABL depth in all models. The largest errors occur for shallow, stably-stratified layers, approach a minimum for weakly convective conditions, and then increase again towards strong convection. In stably-stratified boundary layers we can expect a strong increase in the model departure from observed temperature trends/variability since it is in these shallow boundary layers that the temperature response is most sensitive to the forcing, so a small difference in the depth of the ABL can lead to a large difference in the temperature trend/variability.

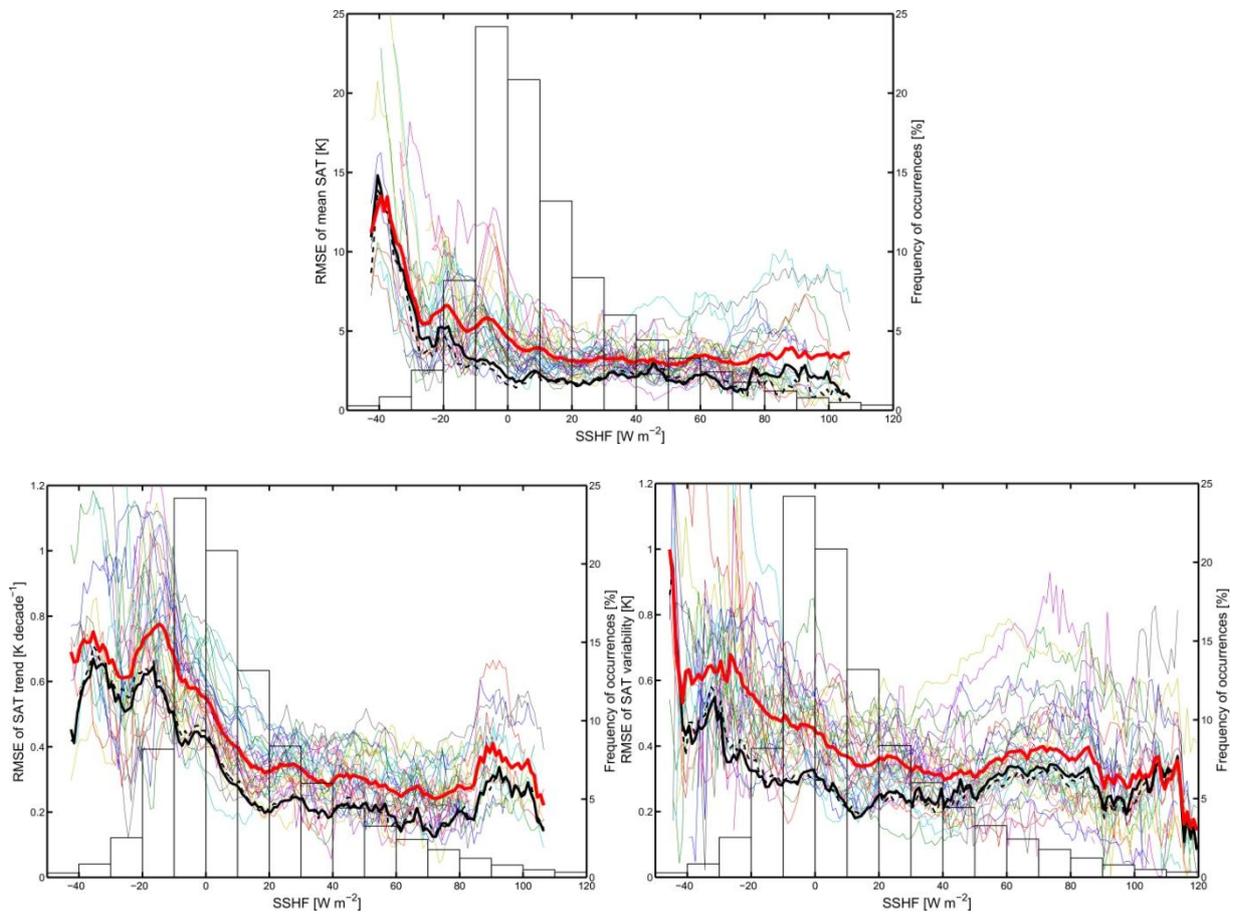

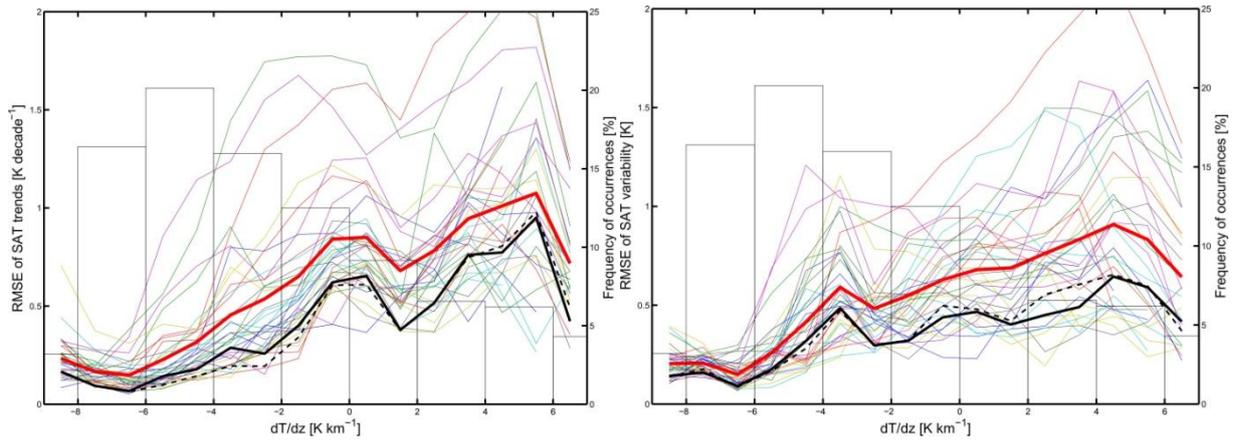

**Figure 2.** The RMS error in the mean, trend and standard deviation in surface air temperature, as a function of the mean surface sensible heat flux over land, and as a function of lower troposphere stability over ocean. The colored lines represent 36 of the model results from the CMIP5 simulations, and the multi-model mean and median are highlighted in solid and dashed black lines respectively. The thick red line is the mean of the model errors in each bin (i.e. the 'typical' model error).

The agreement between the models as to the temperature trends and variability depends upon the state of the ABL. The inter-model mean and spread in the temperature trends and variability, as a function of the state of the ABL over land and ocean is given in Figure 3. There is a large inter-model spread under cold, stably-stratified conditions which approaches a minimum as we move to weakly convective conditions, before increasing again towards deep convection. We see a similar pattern in the inter-model spread for both the temperature trends and variability over land and ocean. The strongest trends are seen in stable conditions, where we also see the greatest spread between the models. The temperature variability can be divided into two regions with a sharp transition zone: in stable conditions there is a high inter-annual variability, and a large inter-model spread, whereas in convective conditions there is a much lower variability, and a smaller spread.

This is to be expected since the shallow ABLs that form in stable stratification amplify temperature changes relative to deep boundary-layers, under a given forcing. So in shallow layers we not only get the stronger temperature response to forcing – the greater trends and variability – but also the different climatology of the ABL between models leads to greater inter-model spread under these conditions. We also see an increasing inter-model spread as we move towards deep convective conditions. Climate models also have trouble representing the ABL depth under strongly convective conditions as many physical processes that occur in these conditions (e.g. self-organization of turbulent structures) are not accounted for in their parameterization schemes. However, due to the reciprocal relationship between ABL depth and temperature response, differences in the ABL depth are less important in determining the SAT trends and variability in deep convective layers, and so the inter-model spread is not as strong as in the shallow, stably-stratified conditions.

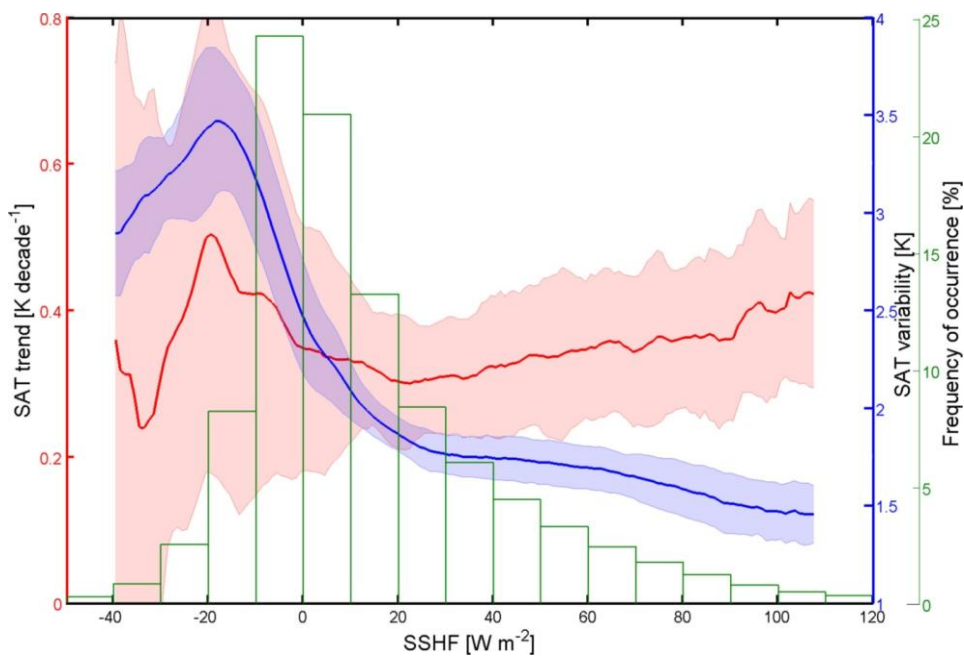

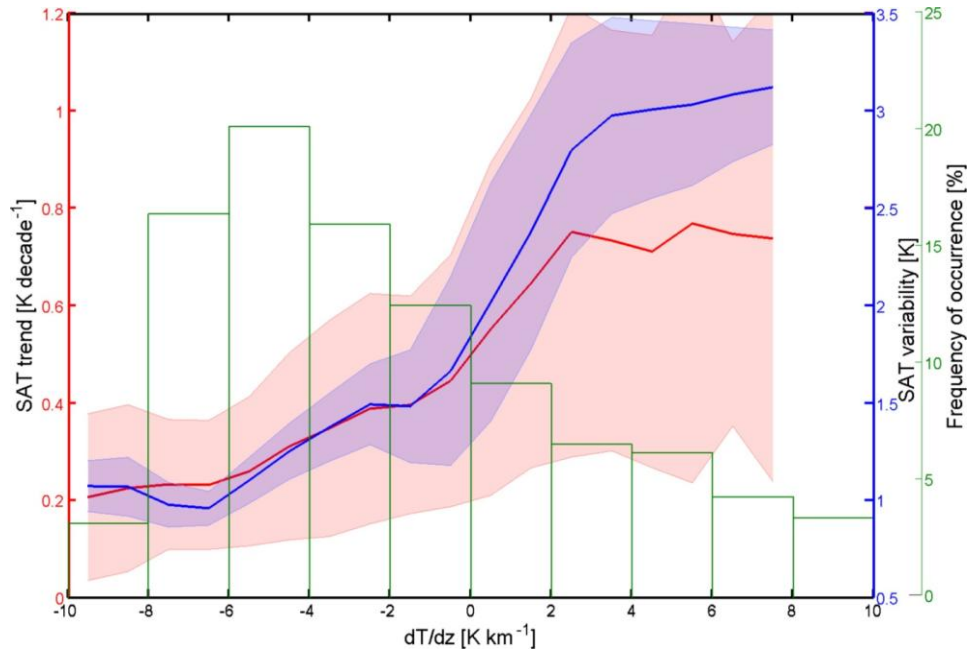

**Figure 3**. The inter-model mean (thick line) and spread (one standard deviation in the model-means, shaded area) in surface air temperature trends (red) and temperature variability (blue) from 36 CMIP5 model results over land (upper figure) and ocean (lower figure). The bar chart indicates the frequency of the occurrence of the surface sensible heat flux for the over-land results and temperature stability for the over-ocean results.

## Quantifying the role of the ABL in introducing model error

The typical geo-spatial pattern of the RMS error in the mean, trend and variability of the surface air temperature is shown in figure 4. These errors represent the degree to which the models describe the climatological annual cycle of these fields at each location. Over ocean we see the greatest model error in mean SAT in the marginal ice zone from the Bering sea to the coast of Greenland. We expect this large departure from observations over the marginal ice zone to be related to differences in the sea-ice extent, since the presence or absence of sea-ice will strongly affect surface heat fluxes, and thus surface air temperature. There are similar hot-spots in the error in variability over the marginal ice zone, but we also see large errors over the continental interior in Asia and North America. These correspond with regions of large departure from observations in the SAT trends. These high-latitude continental interior regions are dominated by very shallow ABLs throughout the winter – and so a poor representation of the ABL in this region would be expected to lead to errors in the depicted annual cycle of SAT trends and variability.

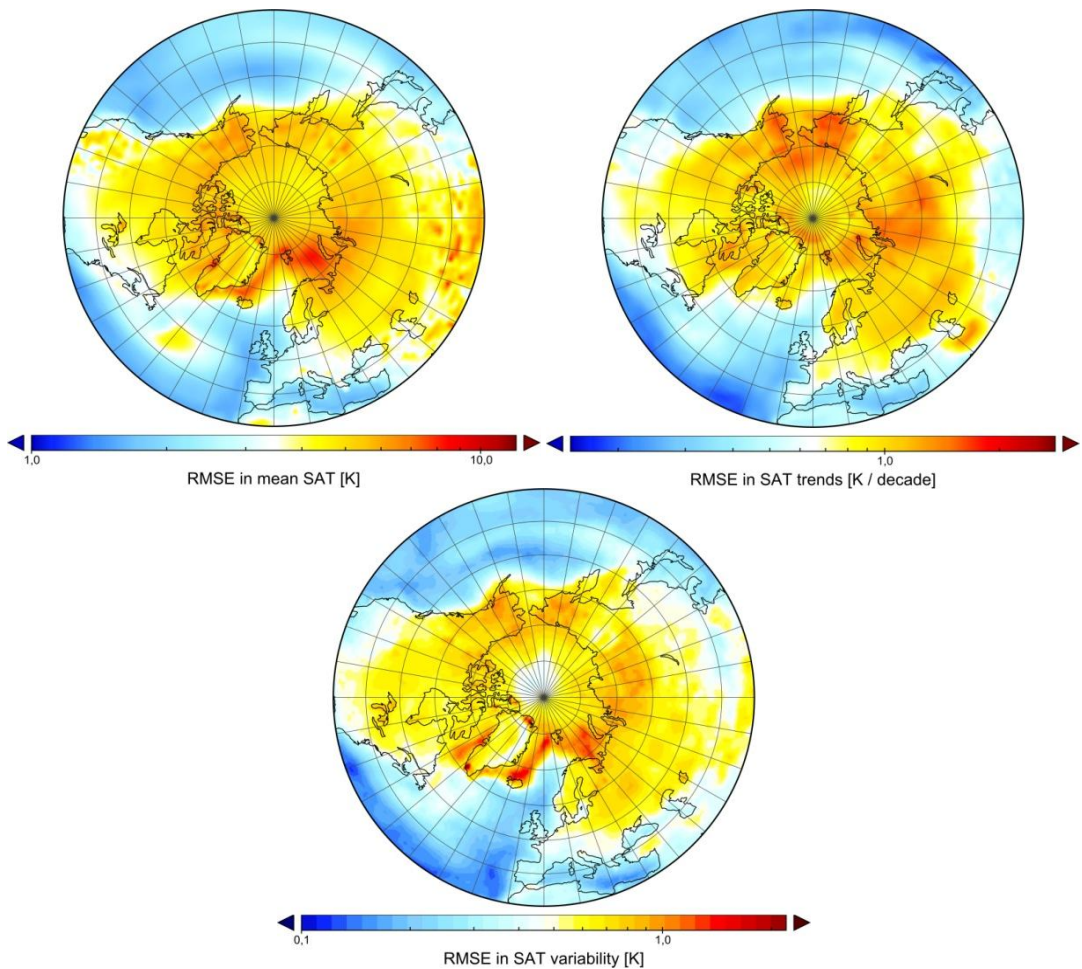

**Figure 4**. The mean of the RMS error in the mean, trend and variability in the surface air temperature for the 36 member ensemble of CMIP5 model results.

Part of the reason the models have a large departure from the observations may be expected to be due to differences in the surface fluxes between the models and observations, especially in the marginal ice zone. We used a multi-linear regression analysis to quantify how much of the geographical pattern of the model error in temperature could be explained by the error in forcing (differences in the fluxes between the model and observations), compared to the bias in the ABL depth. Given that the bias in the ABL depth scales with the mean depth of the ABL, we used the mean surface sensible heat flux as a proxy for bias in the ABL depth over land. The error in the forcing is determined from the RMS error in the surface heat flux between model and reanalysis. The regression coefficients for each model and the 'typical' models (mean and median of model errors) are given in Figure 5, along with the correlation (the quality of fit) between the RMS error from the model and that predicted based on the patterns of bias in ABL depth and error in forcing. There are large differences between the models as to the relative contribution of these components, but in almost all cases the bias in the ABL depth is the better predictor for the error in the temperature metric. In a typical model (Mean Error) the bias in ABL depth can explain up to 55%, 68% and 64% of the geographical variation in the RMS error in temperature mean, trend and variability respectively. In the typical model the RMS error in forcing plays a small or negligible role in determining the pattern of errors in the temperature.

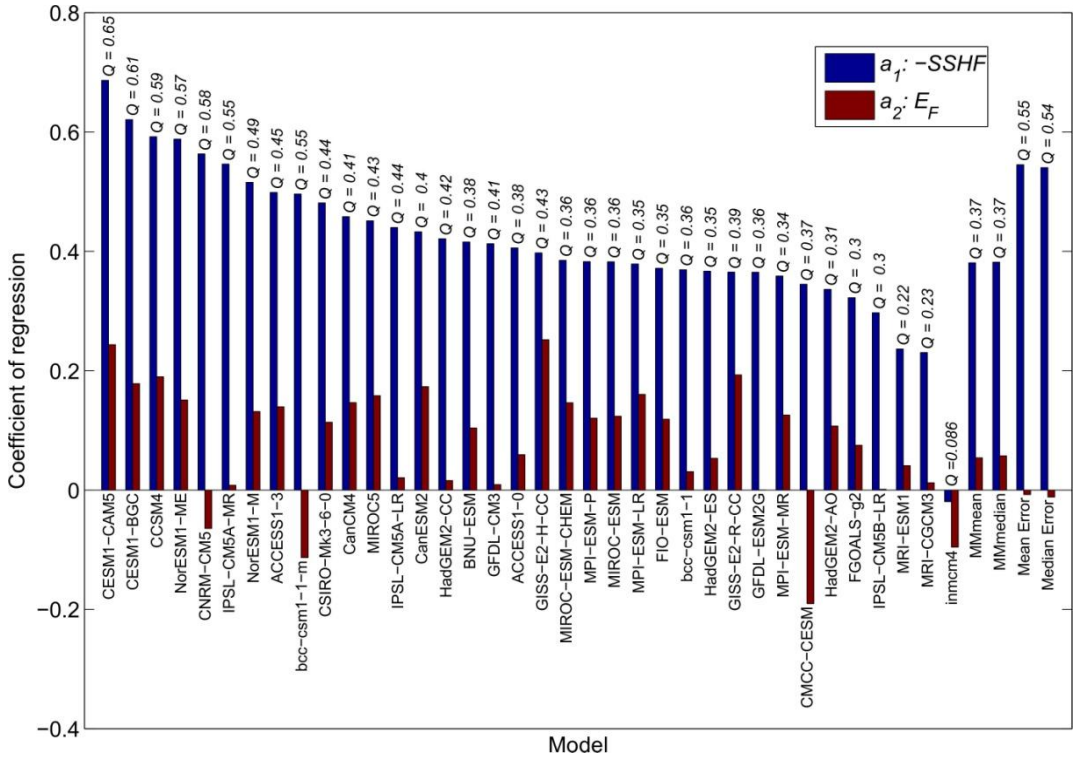

(A) Regression against RMS error in the mean surface air temperature

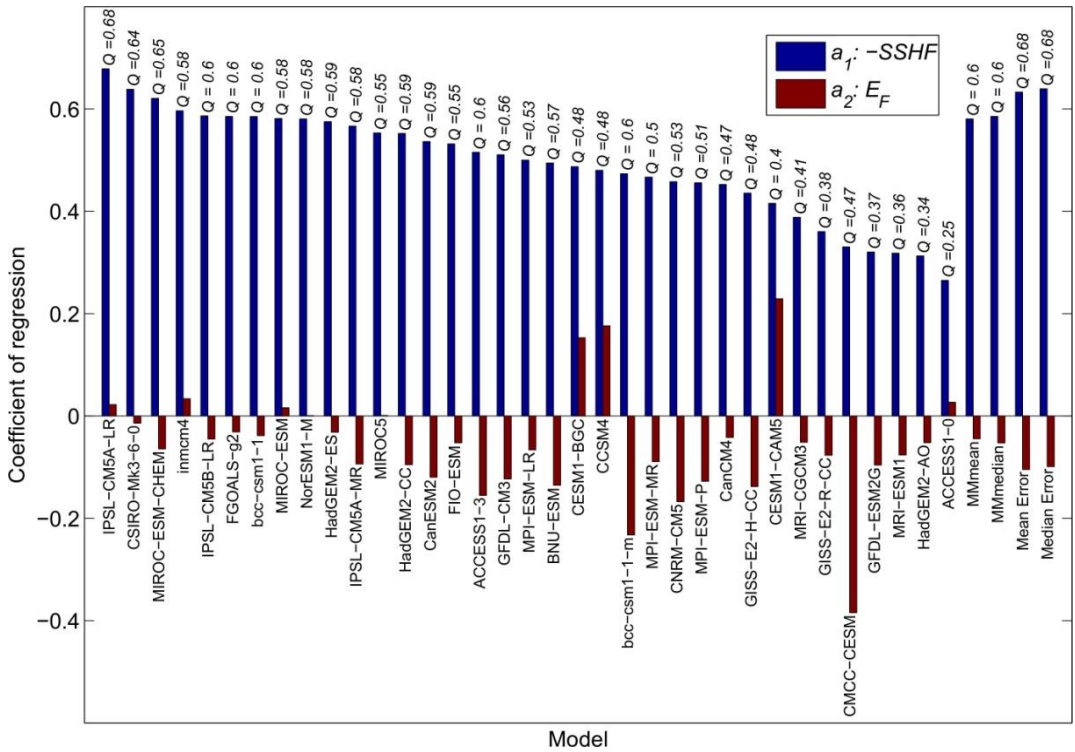

(B) Regression against RMS error in surface air temperature trend

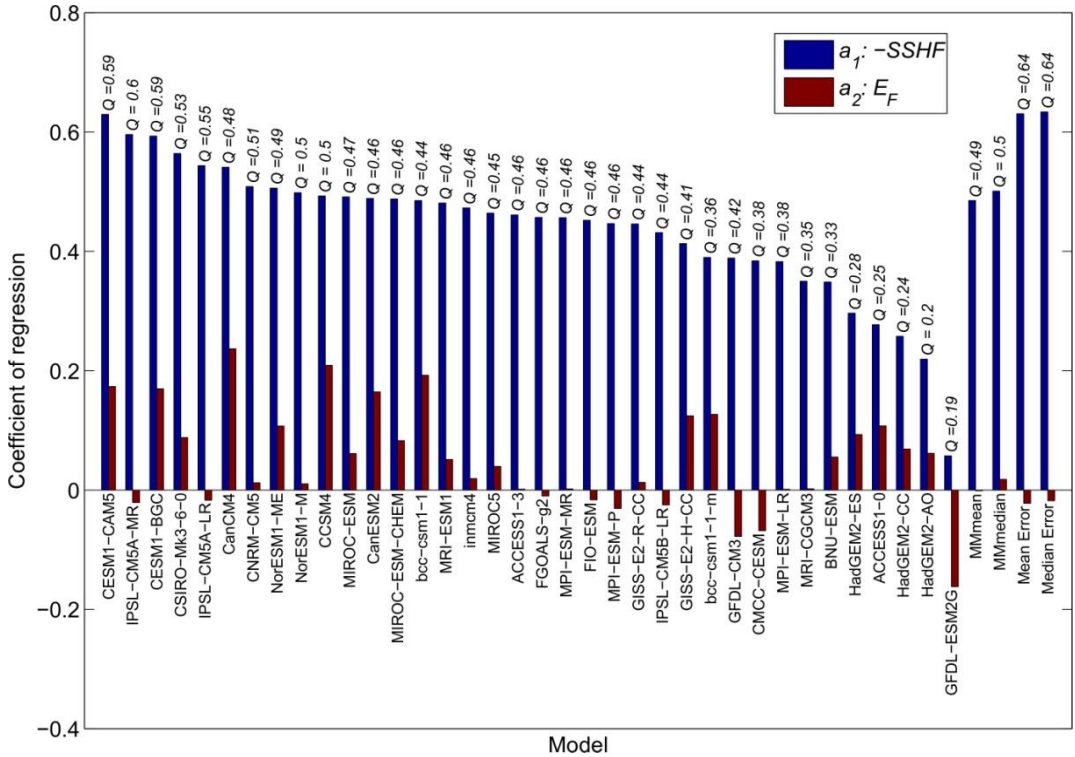

(C) Regression against RMS error in surface air temperature variability

**Figure 5**. The regression coefficients for the mean surface sensible heat flux (blue, $a_1$ from equation 4) and RMS error in surface fluxes (red, $a_2$ from equation 4) for each of the models. The correlation coefficients between the RMS error predicted by the multi-linear regression model and the actual model error are given above the bars. These are given for the multi-linear regression models for the RMS error in (a) mean, (b) trend and (c) variability of the surface air temperature.

In order to test the sensitivity of this analysis to the period under investigation, the analysis was repeated for periods covering the first and last 20 years under investigation: 1979-2000 and 1986-2005. The geospatial pattern of the RMSE in SAT trends is somewhat sensitive to the period under investigation; however, the pattern of RMSE and model biases as a function of boundary-layer conditions is not. It is always conditions the shallow boundary layers that have the greatest

RMSE for the SAT trends. The other metrics showed very little change, neither in the spatial pattern, nor as a function of boundary layer depth. This is not surprising since SAT trends are relatively sensitive to the period selected, compared to the SAT mean and variability.

**Discussion**

There has been a lot of development of the global climate models that have contributed to CMIP since the last phase of the project, CMIP3. However, there are still limitations in the models' ability to describe observed climate variability and trends, especially in cold conditions. Some of these limitations come from constraining the boundary conditions on the model, while others derive from the representation of physical processes in the models – the parameterization schemes.

Our assessment of the CMIP5 model biases and error, with respect to reanalysis, has highlighted the relative poor performance of these models in stably-stratified conditions. It is in these conditions that the models show the greatest departure from observations, and have the greatest differences between the models, in their representation of the surface conditions. This is partly due to the representation of the ABL in global climate models, which are biased towards over-estimating the amount of mixing and producing deeper-than-observed ABLs. This contributes to a bias in the surface conditions whereby the models under-estimate the temperature trends and variability. The problems global climate models have representing the observed variations in the ABL are not surprising given that the parameterization schemes they use are tuned to neutral conditions. These schemes do a reasonably good job of reproducing the observed structure of the typical ABL of mid-latitudes, but miss some essential physics of the turbulence in the ABL under stable stratification and strong convective conditions[33]. This is especially important for shallow, stably-stratified layers which amplify surface temperature response to forcing. Strongly stably-stratified conditions are rare and so even large model error under these conditions may not significantly affect global-mean projections. However, given the importance of the ABL depth in determining the response of the surface temperature to enhanced forcing, this highlights the importance of improving the representation of stably-stratified ABLs in global climate models. A more accurate representation of shallow ABLs would improve the models ability to describe both the SAT mean and variability in the mean climatology, and the trend in surface temperature under an enhanced forcing. This is especially important in the Polar Regions which frequently have a strong surface inversion.

**Supplementary information** is linked to the online version of the paper at www.nature.com/nature.

**Acknowledgments** We thank P. Thorne and S. Outten for critical discussion, and we acknowledge the World Climate Research Programme's Working Group on Coupled Modelling, which is responsible for CMIP, and we thank the climate modeling groups for producing and making available their model output. For CMIP the U.S. Department of Energy's Program for Climate Model Diagnosis and Intercomparison provides coordinating support and led development of software infrastructure in partnership with the Global Organization for Earth System Science Portals. We thank the ECMWF for providing ERA-Interim.

**Author contributions** I.E. and R.D. were responsible for designing the study. R.D. was responsible for obtaining and analyzing the data, and reporting on the findings.

**Author contributions** Reprints and permissions information is available at www.nature.com/reprints. Correspondence and requests for materials should be addressed to R.D. (Richard.davy@nersc.no).

**Data and methods**

Here we analyzed results from the historical simulations of the CMIP5 program. We used 36 of the model results that contributed to the CMIP5 which were created by: CSIRO Australia's CSIRO Mk 3.6[17]; the Japan agency for Marine-Earth Science and Technology and National Institute for Environmental Studies MIROC-ESM(-CHEM)[18] and MIROC5[19]; The Centro Euro-

Mediterraneo sui Cambiamenti Climatici's CMCC-CESM; the Japanese Meteorological Research Institute's MRI-CGCM3 and MRI-ESM1; the Chinese Institute for Atmospheric Physics's FGOALS-g2[20], the French National Centre for Meteorological Research's CNRM-CM5(-2)[21]; the Canadian Earth System Model CanESM2[22] and coupled model CanCM4; the Max-Planck Institute's MPI-ESM-(LR/MR/P); the UK Met office Hadley Centre's HadCM3 and HadGEM2-(AO/CC/ES)[23]; the Institute of Pierre Simon Laplace's IPSL-CM5[24]; the institute for numerical mathematics' INMCM4[25]; the Chinese First Institute of Oceanography's FIO-ESM[26]; the Beijing Normal University's BNU-ESM; the Geophysical Fluid Dynamics Laboratory's GFDL-CM3 and GFDL-ESM2G[27]; NASA's GISS-E2-(H/R)-CC; the Beijing Climate Center's BCC-CSM1(m); the center for Australian Weather and Climate Research's ACCESS 1-0/1-3[28] and the Norwegian Climate Centre's NorESM1-M(E)[29]. For the reference dataset we used the European Center for Medium-range Weather Forecast's (ECMWF's) Interim reanalysis: ERA-Interim. The historical simulations of the CMIP5 program were run from 1850 to 2005, with imposed conditions for the atmospheric composition, solar forcing, natural and anthropogenic aerosols and conditions of land use based on observations[30]. From this ensemble of models we constructed two further datasets by taking the mean and median of the data at each time step to create a Multi-Model Mean and Median model (MMMean and MMMedian respectively).

We use three metrics by which to assess the models' ability to simulate the climate system: the mean, trend and standard deviation of the surface air temperature. These are calculated for each month over the period 1979-2005. The trend is calculated using a least-squares linear regression, and the standard deviation is calculated from the anomalies with respect to the climatological mean temperature for each month. To assess model fidelity with respect to these metrics, we define root-mean square error (E) between a model field (F) and a reference data set (R) such that:

$$E_{i,j}^2 = \sum_t w_t (F_{i,j,t} - R_{i,j,t})^2 \qquad (2)$$

where $i$, $j$ and $t$ are indices for the longitude, latitude and time dimensions respectively, with individual weights, $w_t$, which are determined by the length of each month[15]. Model bias is defined as the mean difference between the model field and the reference data at each location. Note that in order to compare model fields with our reference dataset, the model fields, $F$, were interpolated onto the grid of the reference data, $R$, using a bicubic spline interpolation. We used the above definition for model fidelity to assess the geographical pattern of the model error and biases.

In order to assess the model error as a function of the boundary-layer conditions we use established proxies for the boundary layer depth[31]. Over land we use the surface sensible heat flux, with strong positive heat fluxes indicating deep, convective boundary layers (ABL depth ~ up to few km), and negative values indicating shallow, stably-stratified boundary layers (ABL depth ~ 10-300 m). Over the ocean we use the lower-tropospheric stability ($dT/dz$, defined as the gradient in temperature between the surface and the 850hPa level) as our measure of the boundary-layer conditions, with negative values indicating a warmer surface than upper level, and so relatively deep convective boundary layers, and positive values indicating a colder surface, and a relatively shallow boundary layer.

To get the root mean square error (RMSE) as a function of surface sensible heat flux (SSHF) we took a bin average for each metric as a function of the mean SSHF using a moving window of width 5 Wm$^{-2}$. The RMSE was then calculated for each bin by summing the square of the model error in that bin over all months:

$$E_{SSHF}^2 = (\sum_t w_t (F_t - R_t)^2)_{SSHF} \tag{3}$$

We used the same method to assess each metric as a function of *dT/dz*, using a bin-width of 2 K km$^{-1}$. The model bias is defined as the mean of the difference between the model field and the reference data at each value of the surface sensible heat flux. To account for conditions in which the ABL is exposed to the free atmosphere - and therefore we do not expect to see the same relationship between ABL depth and temperature response as described in equation 1 - we apply a mask to the data to remove all locations where the surface is greater than 1 km above sea level[32]. While this does exclude high-altitude plains locations from our analysis in which we do not expect to have the same problem of ABL exposure to the free atmosphere, such locations are sufficiently few that their exclusion does not significantly alter our results. From the ensemble of model results we constructed two 'typical' model errors by taking the mean and median of the model's errors (labeled 'Mean Error' and 'Median Error' respectively).

A multi-linear regression (MLR) model was constructed to assess the relative contributions of the models' uncertainty in the forcing and their bias in ABL depth in explaining the geographical pattern to the models' error in the surface air temperature mean, trends and variability. It has been demonstrated that the models have systematic biases in the ABL depth[15], and so the magnitude of the bias scales with the depth of the ABL. Given the surface sensible heat flux as a proxy for ABL depth, our regression model takes the form:

$$E_{SAT} = a_1(-SSHF) + a_2 E_F + a_3 \qquad (4)$$

where $E_{SAT}$ is the RMS error in the surface air temperature metric (mean, trends or variability), $SSHF$ is the mean surface sensible heat flux, $E_F$ is the error in the forcing (the RMS error in surface heat flux) and $a_i$ are the coefficients of regression. The predictor and modeled variables were standardised (mean removed and normalised by their standard deviation) so that the coefficients of regression are directly comparable.

The above MLR model gives us the best-fit for the parameters $a_{1,2,3}$ which gives us a method for predicting the pattern of a model's errors in the temperature metric, given the climatological mean surface sensible heat flux and the model's error in the forcing. This takes the same form as our best-fit model above: $\hat{E}_{SAT} = a_1(-SSHF) + a_2 E_F + a_3$, where $\hat{E}_{SAT}$ is the predicted pattern of error in the temperature. The correlation between the RMS error that was found in the model, $E_{SAT}$, and that predicted from the multi-linear regression model, $\hat{E}_{SAT}$, tells us how much of the geographical variation in the model error in temperature can be explained by our MLR model. So the quality of the fit, $Q$, is given by:

$$Q = \frac{n \sum E_{SAT} \hat{E}_{SAT} - (\sum \hat{E}_{SAT})(\sum E_{SAT})}{\sqrt{n \sum E_{SAT}^2 - (\sum E_{SAT})^2} \sqrt{n \sum \hat{E}_{SAT}^2 - (\sum \hat{E}_{SAT})^2}} \tag{5}$$

where n is the number of paired data points.

The models were assessed as an ensemble by considering the mean and variation (inter-model spread) in the surface air temperature trends and variability as a function of the boundary-layer depth. For each model we binned the temperature trend, taken for each month at each location, by the relevant proxy for boundary-layer depth. The temperature trend was normalized by the mean value across all boundary-layer depths, and then scaled by the mean trend found for the Multi-Model mean dataset. This removed any systematic biases in a given model so that we could consider how the model trends varied as a function of boundary layer depth alone. The method was repeated for the temperature variability.